\documentclass[twocolumn,superscriptaddress,aps,pra,preprintnumbers,10pt]{revtex4-2}
\usepackage{graphicx} 
\usepackage{amsmath} 
\usepackage{amssymb} 
\usepackage[T1]{fontenc}
\usepackage{bbold}
\usepackage[urlcolor=blue,colorlinks=true,citecolor=blue,linkcolor=blue,pdfstartview={FitH},bookmarks=false]{hyperref}
\usepackage{enumitem}
\usepackage{xcolor}
\usepackage[utf8]{inputenc}

\date{\today}
\begin{document}
\title{Particle-Hole Asymmetry and Pinball Liquid in a Triangular-Lattice Extended Hubbard Model within Mean-Field Approximation}

\author{Aleksey Alekseev}
\homepage[\mbox{ORCID ID}: ]{https://orcid.org/0000-0001-5102-6647}
\affiliation{\mbox{Institute of Spintronics and Quantum Information, Faculty of Physics and Astronomy}, Adam Mickiewicz University in Pozna\'n, Uniwersytetu Pozna\'nskiego 2, PL-61614 Pozna\'n, Poland}
\author{Agnieszka Cichy}
\homepage[\mbox{ORCID ID}: ]{https://orcid.org/0000-0001-5835-9807}
\affiliation{\mbox{Institute of Spintronics and Quantum Information, Faculty of Physics and Astronomy}, Adam Mickiewicz University in Pozna\'n, Uniwersytetu Pozna\'nskiego 2, PL-61614 Pozna\'n, Poland}
\affiliation{Institut f\"ur Physik, Johannes Gutenberg-Universit\"at Mainz, Staudingerweg 9, D-55099 Mainz, Germany}
\author{Konrad Jerzy Kapcia}
\email[e-mail: ]{konrad.kapcia@amu.edu.pl}
\homepage[\mbox{ORCID ID}: ]{https://orcid.org/0000-0001-8842-1886}
\affiliation{\mbox{Institute of Spintronics and Quantum Information, Faculty of Physics and Astronomy}, Adam Mickiewicz University in Pozna\'n, Uniwersytetu Pozna\'nskiego 2, PL-61614 Pozna\'n, Poland}

\begin{abstract}
Recently, triangular lattice models have received a lot of attention since they can describe a number of strongly-correlated materials that exhibit superconductivity and various magnetic and charge orders. In this research we present an extensive analysis of the charge-ordering phenomenon of the triangular-lattice extended Hubbard model with repulsive onsite and nearest-neighbor interaction, arbitrary charge concentration, and $\sqrt{3}\times\sqrt{3}$ supercell (3-sublattice assumption). The model is solved in the ground state with the mean-field approximation which allowed to identify $8$ charge-ordered phases and a large variety of phase transitions. An exotic pinball-liquid phase was found and described. Moreover, strong particle-hole asymmetry of the phase diagram is found to play an important role for triangular lattices. The detailed analysis of band structures, unavailable for more advanced methods, such as dynamical mean-field theory, allowed us to interpret the found triangular-lattice phases and provided a great insight into the mechanisms behind the phase transitions that can also be met when correlation effects are taken into account. The complexity of the mean-field phase diagram showed the importance and usefulness of the results for the further research with correlation effects included. Together with atomic-limit approximation it can serve them as both a starting point, and a tool to interpret results.
\end{abstract}

\maketitle

\section{Introduction}

A 2D triangular lattice is typical for a number of organic conductors \cite{Tomioka1995, Tokura1996, Takano2001}, transition-metal oxides and dichalcogenides, can be formed by adsorbed helium atoms on a surface, and can describe the moir\'e lattices \cite{wu2018, McDonald2011, Chen2023}. The latter is an interesting platform to investigate various strongly-correlated and frustration-induced phenomena, since the interaction parameters and carrier itineracy can be controlled by changing a twist angle and the choice of two layers out of rich family of 2D materials \cite{Kim2017, Cao2018, Herrero2018}. Another useful platform to investigate the triangular lattice experimentally are the ultracold atomic gases on optical lattices \cite{yang2021, Struck2011, Mathey2007, Yamamoto2020, Mongkolkiattichai2023, Wu2024, Wessel2007, Jo2012}. The systems with triangular-lattice structure are found to host various phenomena, such as superconductivity \cite{Kim2023, Venderley2019, Chen2013, Hong2021, Horigane2019, Yamada2025, Watanabe2005, Cheng2024}, variety of charge and magnetic orderings \cite{Morera2023, Ciorciaro2023, Khatua2022, Sala2023, Ferreira2020, Ogunbumni2022, Xing2020, Isono2020, Yoshioka2009, Shirakawa2017}, topological states \cite{Jiang2020, Szasz2020}. Among them, the charge ordering, e.g., (generalized) Wigner crystals \cite{Noda2002, Hiraki1998, Tan2023, Matty2022}, charge-transfer insulators, charge-density waves \cite{Kuramoto1978, Zarenia2017, Pankov2008, Amaricci2010}, or charge glasses, attracts researchers interest due to its interplay with superconductivity, as well as, possible applications in new devices, such as involving pyroelectric or ferroelectric materials \cite{Mikolajick2021}. Triangular lattices are commonly recognized to be perspective for searching the exotic charge orders, e.g., a pinball-liquid (PL) order \cite{hotta2006,hotta2007, merino2013, Ralko2015, Oles2012, Miyazaki2009}. The PL phase consists of the lattice sites that are insulating for the charge carriers (pins) but surrounded with the lattice sites where the charge carriers are itinerant (balls), and thus the PL shares some properties of supersolids \cite{Rakic2024}.

The model for the investigation of charge-ordering phenomenon is the extended Hubbard model (EHM) \cite{Micnas1990, Rosciszewski2003, Hirsch1984, Lin1986, Clay1999, Penc1994, Calandra2002, Davoudi2006, LitakJMMM2017}. It has been investigated with the mean-field approximation (MFA) both in the context of organic conductors for the quarter-filling or 3/4-filling \cite{seo2000, kaneko2006}, and in the context of the moiré lattices for a series of a few fillings \cite{Tan2023,Ung2023,Pan2020}. The methods beyond the MFA, such as the dynamical mean-field theory \cite{Georges1996, Pietig1999, Tong2004}, are actively used to investigate the triangular-lattice materials with a specific charge concentration as well \cite{Tan2023,merino2013}. 
Nevertheless, to understand the full picture of various charge orders in the triangular lattice, the investigation in the grand-canonical ensemble with arbitrary charge concentration is required. 
It is especially relevant since a particular charge concentration may appear in a phase-separation region: this can be captured as a first-order transition within the fixed-chemical-potential approach, while the fixed-concentration approach might yield an artificial result without consideration of phase separated states explicitly, cf. e.g., \cite{KapciaAPPA2016,KapciaAPPA2018}.
Here we take advantage on the investigation of the EHM itself, encompassing both existing and theoretical materials with arbitrary repulsive electron interaction, in search of attractive properties and the mechanisms behind them. 
It is worth to highlight that a simple Hubbard model provided significant insights into condensed-matter physics and described various phenomena despite its extreme simplicity and artificiality, while the EHM is its most obvious continuation. 
Meanwhile, the successes in the ultracold atomic gases in optical lattices have also renewed the interest in Hubbard-like models, since the physics of such quantum simulators is not just modeled with these models, but described by them.

The mean-field study of the triangular-lattice EHM in the grand-canonical ensemble can provide significant insights into the problem despite ignoring the correlation effects. 
One should remember that the Hartree-Fock MFA overestimates critical temperatures and particularly the temperature range of stability of long-range orders, but it can, nevertheless, give a qualitative description of the system in the ground state in certain interaction parameter ranges \cite{Micnas1990,KapciaAPPA2016,KapciaAPPA2018}.
Besides using it as a benchmark for further investigations, a number of unusual phenomena can already be found within the MFA, which makes them both easy to analyze and distinguish with strongly-correlated phenomena \cite{Kapcia2011, Kapcia2016}. The great advantage here (besides the required computational and time resources) is provided by the opportunity to analyze band structures in details, which cannot be done within dynamical mean-field theory that we consider to be the next step of this research (to take into account local correlations). Moreover, the non-correlated phases are found within the dynamical mean-field theory when the intersite interaction prevails over the onsite interaction \cite{kapcia2017}, while some other phases and phase transitions can have similar qualitative features as withing the MFA. It is advantageous to use MFA results as a reference point in future studies together with the atomic limit results \cite{kapcia2021,KapciaJMMM2022}. 

Here, we present the solution of the triangular-lattice EHM within the MFA focusing on the zero-temperature systems without a magnetic order and utilizing the $\sqrt{3}\times\sqrt{3}$ hexagonal supercells. 
The found phase diagram consists of the large variety of phase transitions and is more complex than one could assume in the absence of strong-correlation effects.
Hence, the results included in this work are i) required as a starting point for the further investigations that would yield even more features in the phase diagram together with being more computationally expensive and suffering from convergence problems; and ii) useful as a tool to interpret the future results while being an extensive standalone research itself.
Both the pinball-liquid phase and the (often ignored) strong particle-hole asymmetry are found and analyzed.

In the paper, after the discussion of the method of investigation (Sec. \ref{sec:method}), we present the known band structures of the non-interacting triangular and honeycomb lattices, required for the further discussion (Sec. \ref{sec:nointeract}), and present the found mean-field phase diagram of the triangular lattice (Sec. \ref{sec:phase}) with a description of the found phases and phase transitions (Secs. \ref{sec:AAB}-\ref{sec:symbreak}). Sec. \ref{atlim} is devoted to the brief comparison of the mean-field and the atomic-limit results. 
The main finding are summarized in Sec. \ref{summary}.

\section{Model and Method}\label{sec:method}

The extended Hubbard model in a grand-canonical ensemble of electrons is represented by the Hamiltonian \cite{Micnas1990, Rosciszewski2003, Hirsch1984, Lin1986, Clay1999, Penc1994, Calandra2002, Davoudi2006, LitakJMMM2017}:
\begin{eqnarray}\label{ehm}
    \hat{H} 
    & = & - t \sum_{\left< i,j \right>, \sigma} \left( \hat{c}^\dag_{i\sigma} \hat{c}_{j\sigma} + \text{H.c.} \right)
    + U\sum_i \hat{n}_{i\uparrow}\hat{n}_{i\downarrow} \\
    & + & V \sum_{\left< i,j \right>} \hat{n}_i \hat{n}_j
    - \mu\hat{N}, \nonumber
\end{eqnarray}
where $t$, $\mu$, $U$, and $V$ are a hopping amplitude, a chemical potential, an onsite and an intersite nearest-neighbor (NN) interaction parameters, respectively. These parameters are effective, meaning they can include not only Coulomb repulsion but other interactions and renormalizations such as those that involve phonons. The $i$ and $\sigma$ are site and spin indices while the summation over $\left<i,j\right>$ means a summation over NN pairs without repeating (i.e., if the term with $ij$ is present in a sum, the term with $ji$ is not). The $\hat{c}^\dag_{i\sigma}$ and $\hat{c}_{i\sigma}$ are creation and annihilation operators, $\hat{n}_{i\sigma} = \hat{c}^\dag_{i\sigma}\hat{c}_{i\sigma}$ is an occupation number operator, $\hat{n}_i = \hat{n}_{i\uparrow} + \hat{n}_{i\downarrow}$, and $\hat{N} = \sum_{i} \hat{n}_i$.

The mean-field approximation
\begin{equation}
    \hat{n}_{i\sigma}\hat{n}_{j\sigma'} = n_{j\sigma'} \hat{n}_{i\sigma} + n_{i\sigma} \hat{n}_{j\sigma'} - n_{i\sigma} n_{j\sigma'}
\end{equation}
($n_{i\sigma} = \left< \hat{n}_{i\sigma} \right>$), and the Fourier transform to a reciprocal space turn the Hamiltonian into a sum of independent terms. In particular, for the triangular lattice and $\sqrt{3}\times\sqrt{3}$ supercell
\begin{eqnarray}\label{hamiltonian}
    \hat{H} &=& \sum_{\mathbf{k}\sigma} \left[ \varepsilon_\mathbf{k}
    (\hat{c}^\dag_{1\mathbf{k}\sigma}\hat{c}_{2\mathbf{k}\sigma}
    + \hat{c}^\dag_{3\mathbf{k}\sigma}\hat{c}_{1\mathbf{k}\sigma}
    + \hat{c}^\dag_{2\mathbf{k}\sigma}\hat{c}_{3\mathbf{k}\sigma})
    +\text{H.c.}+ \right. \nonumber\\
    &+& \left. \sum_\alpha \epsilon_{\alpha\sigma} \hat{n}_{\alpha\mathbf{k}\sigma}\right]+C=\sum_{\mathbf{k}\sigma} \hat{H}_{\mathbf{k}\sigma} + C,
\end{eqnarray}
where $\alpha = 1, 2, 3$ is a sublattice index, $\mathbf{k}$ is a reciprocal-space vector, the constant term
\begin{equation}
    C = - \frac{L}{3} \sum_\alpha \left( 
    U n_{\alpha\uparrow}n_{\alpha\downarrow}
    + \frac{zV}{2} n_{\bar\alpha} n_{\bar{\bar\alpha}}
    \right),
\end{equation}
($\bar\alpha$ and $\bar{\bar\alpha}$ are sublattice indices different from $\alpha$ and from each other), $L$ is the number of lattice sites, $n_\alpha = \frac{3}{L}\sum_{\sigma\mathbf{k}} \left< \hat{n}_{\alpha\mathbf{k}\sigma} \right>$, $z=6$ is a coordination number, 
\begin{equation}\label{en-sbl}
    \epsilon_{\alpha\sigma} = U n_{\alpha\bar\sigma}
    + \frac{zV}{2} (n_{\bar\alpha} + n_{\bar{\bar\alpha}})
    -\mu,
\end{equation}
($\bar\sigma$ is the spin index different from $\sigma$)
and
\begin{equation}
    \varepsilon_\mathbf{k} = - t \left(
    e^{i\mathbf{k}\mathbf{r}_1} + e^{i\mathbf{k}\mathbf{r}_2} + e^{i\mathbf{k}\mathbf{r}_3}
    \right)
\end{equation}
with:
\begin{equation}
    \mathbf{r}_1 = \left(-\frac{1}{3},-\frac{2}{3}\right), \
    \mathbf{r}_2 = \left(-\frac{1}{3},\frac{1}{3}\right), \
    \mathbf{r}_3 = \left(\frac{2}{3},\frac{1}{3}\right),
\end{equation}
provided that the vectors $\mathbf{k}$ are written in the basis of a cell that is reciprocal to the $\sqrt{3}\times\sqrt{3}$ supercell.

Each term $\hat{H}_{\mathbf{k}\sigma}$ in the Hamiltonian (\ref{hamiltonian}) can be written as 
\begin{equation*}
    \hat{H}_{\mathbf{k}\sigma} = 
    \hat{I}_{\mathbf{k}_1\uparrow} \otimes \hat{I}_{\mathbf{k}_1\downarrow} \otimes \hat{I}_{\mathbf{k}_2\uparrow}
    \otimes \ldots \otimes
    \hat{\mathbf{H}}_{\mathbf{k}\sigma} 
    \otimes \ldots \otimes
    \hat{I}_{\mathbf{k}_{L/3}\downarrow},
\end{equation*}
where $\hat{\mathbf{H}}_{\mathbf{k}\sigma}$ can be represented by in a form of a block-diagonal $8 \times 8$ matrix:
\begin{equation}\label{hammatrix}
    \hat{\mathbf{H}}_{\mathbf{k}\sigma} = \begin{pmatrix}
        0 & \mathbb{0}_{1\times 3} & \mathbb{0}_{1\times3} & 0 \\
        \mathbb{0}_{3\times1} & \check{\mathbb{H}}_{1\mathbf{k}\sigma} & \mathbb{0}_{3\times3} & \mathbb{0}_{3\times1}\\  
        \mathbb{0}_{3\times1} & \mathbb{0}_{3\times3} & \check{\mathbb{H}}_{2\mathbf{k}\sigma} & \mathbb{0}_{3\times1}\\
        0 & \mathbb{0}_{1\times3} & \mathbb{0}_{1\times3} & \epsilon_1 + \epsilon_2 + \epsilon_3
    \end{pmatrix}_{\mathbf{k}\sigma},
\end{equation}
Here, $\mathbb{0}_{m \times n}$ denotes a block of $m \times n$ size  with all elements $0$, whereas $\check{\mathbb{H}}_{1\mathbf{k}\sigma}$ and $\check{\mathbb{H}}_{2\mathbf{k}\sigma}$ are the following $3 \times 3$ matrices:
\begin{equation}\label{eq.bandstructure}
    \check{\mathbb{H}}_{1\mathbf{k}\sigma} \equiv \begin{pmatrix}
    \epsilon_{1\sigma} & \varepsilon_\mathbf{k} & \varepsilon_\mathbf{k}^*\\
    \varepsilon_\mathbf{k}^* & \epsilon_{2\sigma} & \varepsilon_\mathbf{k}\\
    \varepsilon_\mathbf{k} & \varepsilon_\mathbf{k}^* & \epsilon_{3\sigma}
\end{pmatrix},
\end{equation}
\begin{equation}
    \check{\mathbb{H}}_{2\mathbf{k}\sigma} \equiv \begin{pmatrix}
    \epsilon_{1\sigma}+\epsilon_{2\sigma} & \varepsilon_\mathbf{k} & -\varepsilon_\mathbf{k}^*\\
    \varepsilon_\mathbf{k}^* & \epsilon_{1\sigma}+\epsilon_{3\sigma} & \varepsilon_\mathbf{k}\\
    - \varepsilon_\mathbf{k} & \varepsilon_\mathbf{k}^* & \epsilon_{2\sigma}+\epsilon_{3\sigma}
\end{pmatrix}.
\end{equation}
The operator matrix (\ref{hammatrix}) is built from the matrix elements with respect to the basis states
\begin{equation}\begin{gathered}
    \left| 0_{1\mathbf{k}\sigma} 0_{2\mathbf{k}\sigma} 0_{3\mathbf{k}\sigma} \right>, \\
    \begin{aligned}
    \left| 1_{1\mathbf{k}\sigma} 0_{2\mathbf{k}\sigma} 0_{3\mathbf{k}\sigma} \right>, &&
    \left| 0_{1\mathbf{k}\sigma} 1_{2\mathbf{k}\sigma} 0_{3\mathbf{k}\sigma} \right>, &&
    \left| 0_{1\mathbf{k}\sigma} 0_{2\mathbf{k}\sigma} 1_{3\mathbf{k}\sigma} \right>,
    \end{aligned} \\
    \begin{aligned}
    \left| 1_{1\mathbf{k}\sigma} 1_{2\mathbf{k}\sigma} 0_{3\mathbf{k}\sigma} \right>, &&
    \left| 1_{1\mathbf{k}\sigma} 0_{2\mathbf{k}\sigma} 1_{3\mathbf{k}\sigma} \right>, &&
    \left| 0_{1\mathbf{k}\sigma} 1_{2\mathbf{k}\sigma} 1_{3\mathbf{k}\sigma} \right>,
    \end{aligned} \\
    \left| 1_{1\mathbf{k}\sigma} 1_{2\mathbf{k}\sigma} 1_{3\mathbf{k}\sigma} \right>,
\end{gathered}\end{equation}
and the subscript $\mathbf{k}\sigma$ under the matrix (\ref{hammatrix}) means that it is associated with only $\mathbf{k}\sigma$ single-particle states.

Thus, the model can be easily solved on a fine grid of $\mathbf{k}$-vectors when the parameters $V/t$, $U/t$, $\mu/t$, and $n_{\alpha\sigma}$ are provided. 
To distinguish stable and metastable phases the grand potential is calculated as
\begin{equation}\label{eq.omega}
    \frac{\Omega}{L} = \frac{1}{L} \left( C - \frac{1}{\beta}\ln\mathcal{Z} \right)
    = \frac{1}{L} \left( C - \frac{1}{\beta} \sum_{\mathbf{k}\sigma} \ln\mathcal{Z}_{\mathbf{k}\sigma} \right), 
\end{equation}
where $\mathcal{Z} = \prod_{\mathbf{k}\sigma}\mathcal{Z}_{\mathbf{k}\sigma}$ is the partition function of the system, $\mathcal{Z}_{\mathbf{k}\sigma}=\sum_m e^{-\beta E^{(m)}_{\mathbf{k}\sigma}}$, $E^{(m)}_{\mathbf{k}\sigma}$ is the $m$-th eigenvalue of $\hat{\mathbf{H}}_{\mathbf{k}\sigma}$ (i.e., of the matrix (\ref{hammatrix})), and $\beta = T^{-1}$ is the inverse temperature.

The solution of the model gives the occupation numbers:
\begin{equation}\label{eq.occup}
    n_{\alpha\sigma} = \frac{3}{L} \sum_\mathbf{k} \frac{\sum_m n_{\alpha\mathbf{k}\sigma}^{(m)} e^{-\beta E^{(m)}_{\mathbf{k}\sigma}}}{\mathcal{Z}_{\mathbf{k}\sigma}},
\end{equation}
where
\begin{eqnarray}
    n_{\alpha\mathbf{k}\sigma}^{(m)} & = & 
    \left|a_{\alpha\mathbf{k}\sigma}^{(m)}\right|^2
    + \left|a_{\alpha\mathbf{k}\sigma,\bar\alpha\mathbf{k}\sigma}^{(m)}\right|^2 \\
    & + & \left|a_{\alpha\mathbf{k}\sigma,\bar{\bar\alpha}\mathbf{k}\sigma}^{(m)}\right|^2
    + \left|a_{1\mathbf{k}\sigma,2\mathbf{k}\sigma,3\mathbf{k}\sigma}^{(m)}\right|^2, \nonumber
\end{eqnarray}
and $a^{(m)}$ are components of the $m$th eigenvector of the matrix (\ref{hammatrix}).

In this research, we focus on the charge-order phenomenon neglecting possible spin-order formation.
Hence, the equations are simplified such that $n_{\alpha\sigma} = n_\alpha/2$ and $\epsilon_\alpha \equiv \epsilon_{\alpha\sigma}$.

Since the $n_\alpha$ take role of both input and output quantities of the model, it can be solved self-consistently. 
In this research, our point of interest is the zero-temperature phase diagram. However, due to the lack of convergence, the calculations are performed for $T = 10^{-3}\cdot 4.5t$ which stabilizes the algorithm. Still, rather strong mixing is used: when the Fermi level is in a proximity of a singularity of a spectral function, only $0.005-0.07$ fraction of a new solution is used for the next iteration. A strict criterion of convergence is used: $10^{-8}$ for $n_\alpha$ together with $10^{-8}\cdot 4.5t$ for $\Omega/L$. The grid of $\mathbf{k}$-points contained $96 \times 96$ points (817 irreducible points; the $\mathbf{k}$-dependent quantities exhibit the symmetry of the non-symmetry-broken triangular lattice (wallpaper group p6m), despite the fact that the supercell has a lower symmetry (p3m1)).

For the sake of better comparison with other lattice models, the quantities are expressed in the units of a half-bandwidth of a non-interacting triangular lattice $D=4.5t$. For the same reason, the intersite interaction $V$ is expressed in the units of $D/z = 0.75t$. In the plots, the displaced chemical potential $\bar\mu = \mu - U/2 - zV$ is used. 

We denote the total concentration $n = \sum_\alpha n_\alpha / 3$. For the analysis of discontinuous phase transitions, the contributions to the grand potential $\Omega$ (all per lattice site $L$) are defined as the thermal averages of the corresponding terms in the Hamiltonian (\ref{ehm}): kinetic energy ($t$-term contribution), on-site interaction energy ($U$-term contribution), intersite-interaction energy ($V$-term contribution), potential energy ($V$- and $U$-term contributions together), and chemical energy ($\mu$-term contribution). At finite temperatures $\Omega$ also has a contribution $-TS$ ($S$ is an entropy of the system) which is zero in this research.

One of the advantages of MFA is the ability to plot and analyze a band structure $\omega(\mathbf{k})$. 
The Hamiltonian (\ref{hamiltonian}) can be written in the form of
\begin{equation}
    \hat{H} = \sum_{\mathbf{k}\sigma} 
    \begin{pmatrix} \hat{c}^\dag_{1\mathbf{k}\sigma} & \hat{c}^\dag_{2\mathbf{k}\sigma} & \hat{c}^\dag_{3\mathbf{k}\sigma} \end{pmatrix}
    \check{\mathbb{H}}_{1\mathbf{k}\sigma} \begin{pmatrix} \hat{c}_{1\mathbf{k}\sigma} \\ \hat{c}_{2\mathbf{k}\sigma} \\ \hat{c}_{3\mathbf{k}\sigma} \end{pmatrix} + C,
\end{equation}
and the band structure is found as eigenvalues of the matrix $\check{\mathbb{H}}_{1\mathbf{k}\sigma}$ (i.e., the matrix (\ref{eq.bandstructure})).
Note that 
$\mathcal{Z}_{\mathbf{k}\sigma}=\prod_m \left( 1 + e^{-\beta \lambda^{(m)}_{\mathbf{k}\sigma}} \right)$, 
where $\lambda^{(m)}_{\mathbf{k}\sigma}$ is the $m$-th eigenvalue of $\check{\mathbb{H}}_{1\mathbf{k}\sigma}$. 
Such a formulation is in the language of non-interacting quasiparticles, it gives exactly the same results as the reasoning leading to the equations (\ref{eq.omega}) and~(\ref{eq.occup}).

The spectral function is calculated as
\begin{equation}
    A_\alpha(\omega + i\eta) = -\frac{1}{\pi}\frac{3}{L}\sum_\mathbf{k} \operatorname{Im} G_{\alpha\alpha\mathbf{k}}(\omega + i\eta),
\end{equation}
where $\eta=0.001D-0.005D$, and $G_{\alpha\alpha\mathbf{k}}(z)$ are diagonal elements of
a $3 \times 3$ matrix $\left( z\check{I} - \check{\mathbb{H}}_{1\mathbf{k}\sigma} \right)^{-1}$.
When the spectral function is calculated with such a small $\eta$, the grid of $396 \times 396$ $\mathbf{k}$-points or more is used to make plots of spectral function clear.

The calculations are implemented in the Python language with the use of Matplotlib library for visualization.

\section{Non-Interacting Limit}\label{sec:nointeract}

\begin{figure}[ht]
    \centering
    \includegraphics[width=1.0\linewidth]{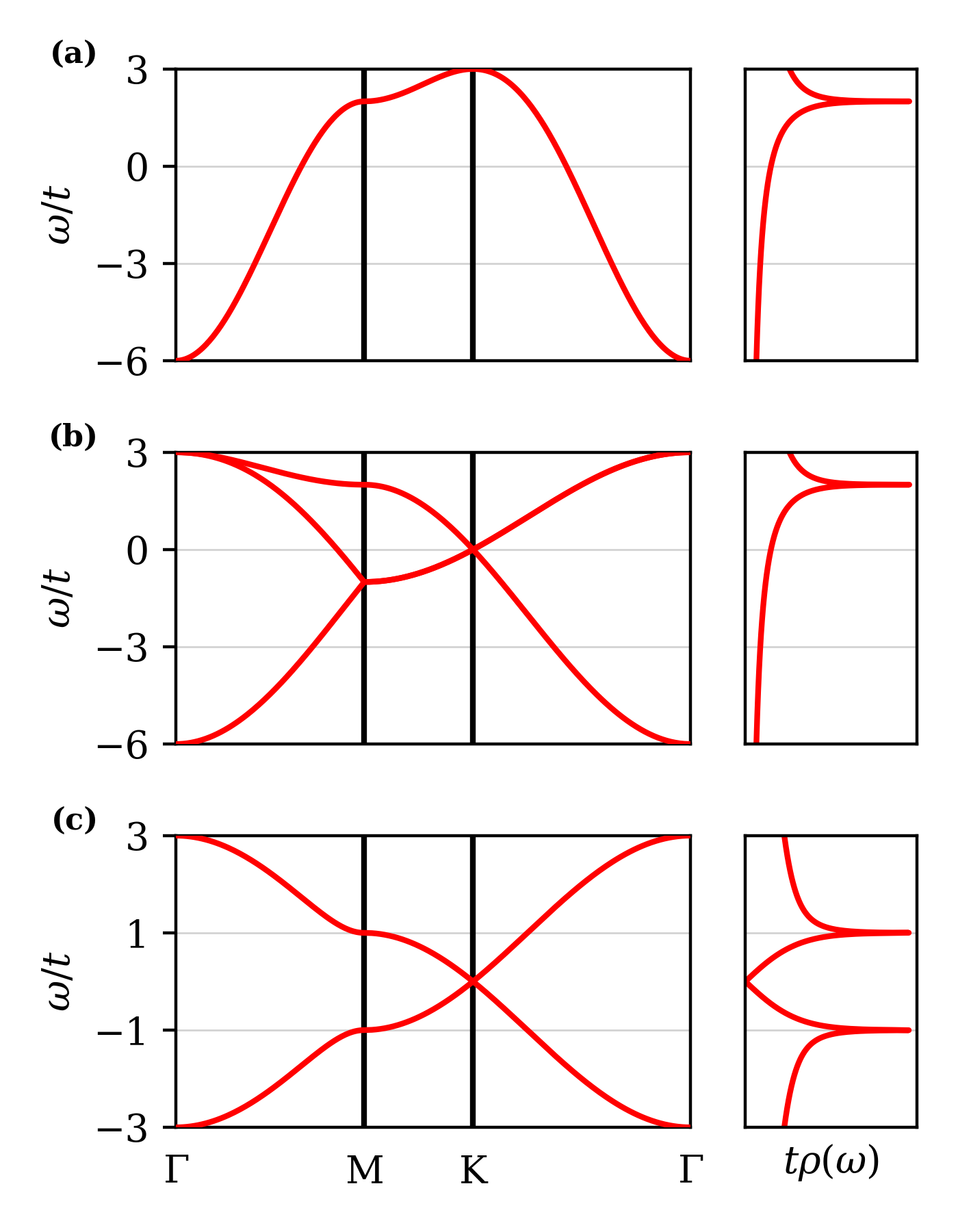}
    \caption{The non-interacting band structures (left) and densities of states (right) of (a) triangular, (b) triangular with the $\sqrt{3}\times\sqrt{3}$ supercell, and (c) honeycomb lattices.}
    \label{fig:non-interacting}
\end{figure}

For the reference point and the following analysis, the band structures and the densities of states (DOSs) of non-interacting triangular and honeycomb lattices are presented in Fig. \ref{fig:non-interacting}. Both the no-sublattice case (Fig. \ref{fig:non-interacting}a) and the 3-sublattice case with the bands folded in accordance with $\sqrt{3}\times\sqrt{3}$ supercell (Fig. \ref{fig:non-interacting}b) are shown.

The DOS (per spin) of a triangular lattice is \cite{hanisch1997}:
\begin{equation}
    \rho_\text{tri}(\omega) = \frac{K(z_1/z_0)}{\pi^2t\sqrt{z_0}},
\end{equation}
where: $K(m)$ is the complete elliptic integral of the first kind,
\begin{eqnarray}
    & & K(m)  =  \int_0^{\frac{\pi}{2}} \frac{d\theta}{\sqrt{1 - m\sin^2\theta}}, \\
    & & z_0  = \begin{cases}
        3 + 2\sqrt{3 - \omega/t} - \left(\frac{\omega/t}{2}\right)^2, &\text{for } 2 \leq \omega/t \leq 3 \\
        4\sqrt{3 - \omega/t}, &\text{for } -6 \leq \omega/t \leq 2
    \end{cases}, \nonumber \\
    & & z_1  =  \begin{cases}
        4\sqrt{3 - \omega/t}, &\text{for } 2 \leq \omega/t \leq 3 \\
        3 + 2\sqrt{3 - \omega/t} - \left(\frac{\omega/t}{2}\right)^2, &\text{for } -6 \leq \omega/t \leq 2
    \end{cases}.  \nonumber
\end{eqnarray}
It is asymmetric, exists from $-6t$ to $3t$, and has a van Hove singularity at $2t$. The zero-temperature half-filling would correspond to a Fermi level $\omega_\text{F} \approx 0.8347t$.

The DOS (per spin) of a honeycomb lattice is \cite{kogan2020,Cichy2022}:
\begin{equation}
    \rho_\text{hcb}(\omega) = 2 |\omega| \rho_\text{tri}(3 - \omega^2).
\end{equation}
It is symmetric around the Dirac cone at $\omega=0t$, exists from $-3t$ to $3t$, and has van Hove singularities at $\pm t$ (Fig. \ref{fig:non-interacting}c).

\section{Phase Diagram}\label{sec:phase}

\begin{figure}[ht]
    \centering
    \includegraphics[width=1.0\linewidth]{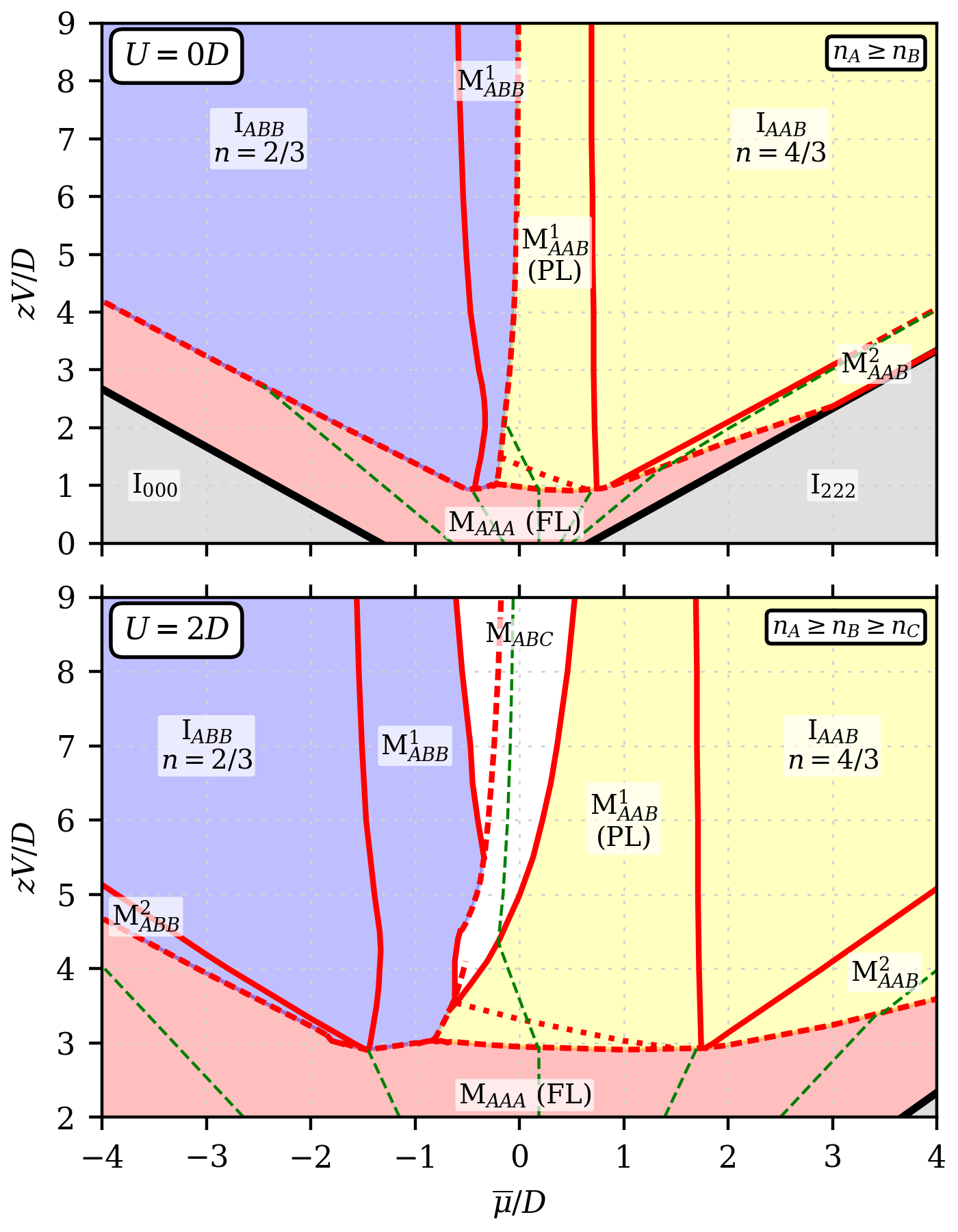}
    \caption{The MFA phase diagram of the triangular-lattice EHM with 3-sublattices as a function of $V$ and $\bar\mu = \mu - U/2 - zV$ for $U=0D$ and $U=2D$. 
    Only the phases with the smallest grand potential are shown. 
    The subscripts in the phase names refer to the charge order (see Fig. \ref{fig:structures}) while letters M and I stand for the metallic and insulating phase, respectively.
    The solid and dashed lines represent continuous and discontinuous phase transitions, respectively. The dotted line is placed at the approximate transition where the charge-ordered metal is not a pinball liquid (PL) anymore. 
    The total concentration is constant along the green dashed lines, in particular $n=$ 1/3, 2/3, 1, 4/3, and 5/3 (from left to right).}
    \label{fig:phase-diagram}
\end{figure}

\begin{figure}[ht]
    \centering
    \includegraphics[width=1.0\linewidth]{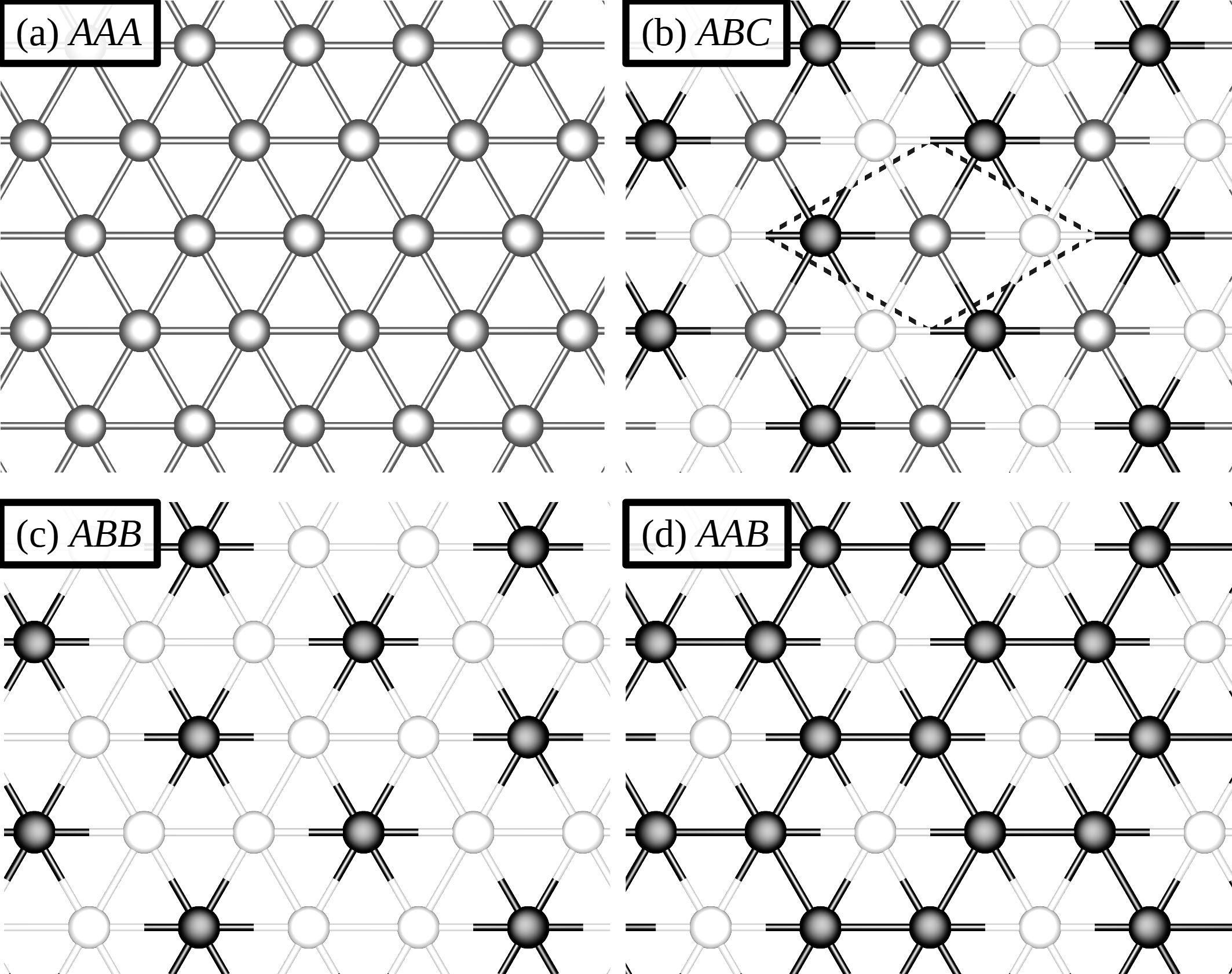}
    \caption{The found charge order types on the triangular lattice. The color of the circles represent the relative charge concentration on the lattice sites: from the highest (black circles) to the lowest (white circles). The $\sqrt{3}\times\sqrt{3}$ supercell is shown by the dashed lines. The figure is prepared using {\sc VESTA program \cite{Vesta}}.}
    \label{fig:structures}
\end{figure}

The phase diagram is shown in Fig. \ref{fig:phase-diagram} while the structures of the found charge orders are schematically presented in Fig. \ref{fig:structures}. 
In the model (\ref{ehm}), the lattice is fully unoccupied (I$_{000}$) for 
$\bar\mu < -(zV + U/2 + 6t)$ and fully occupied (I$_{222}$) for 
$\bar\mu > zV + U/2 + 3t$. The corresponding regions on the phase diagrams are indicated by gray color in Fig. \ref{fig:phase-diagram}. 

The metallic phase without charge order (CO) is denoted as M$_{AAA}$ and marked by the red color. Its band structure and spectral function fully repeat the non-interacting ones (Fig. \ref{fig:non-interacting}b) except the Fermi level is shifted by: 
\begin{equation}
    \omega_\text{shift} = \left(zV+\frac{U}{2}\right)(n - 1) - \bar\mu,
\end{equation}
(cf. Eq. (\ref{en-sbl}) when all $n_{\alpha\sigma}$ are equal).
It is possible to have convergence of an algorithm for the non-charge-ordered phase throughout the whole phase diagram. However, the further it goes into the charge-ordered regions of the phase diagram, the easier it can be destabilized toward CO phases by introduction of small deviations in the input $n_\alpha$, and the larger the difference between its grand potential and the grand potential of the CO phases.

The blue and yellow regions of the phase diagram in Fig. \ref{fig:phase-diagram} mark the CO phases with the sublattice occupation numbers $n_1=n_A$, $n_2=n_3=n_B$ ($ABB$ phases, Fig. \ref{fig:structures}c) and $n_1=n_2=n_A$, $n_3=n_B$ ($AAB$ phases, Fig. \ref{fig:structures}d), respectively, with $n_A > n_B$. These phases have degeneracy of 3, e.g., the possible sublattice occupation numbers $n_\alpha$ of the yellow region ($AAB$ region) are $(n_A, n_A, n_B)$, $(n_A, n_B, n_A)$, and $(n_B, n_A, n_A)$. 

The yellow region (subsection \ref{sec:AAB}) encompasses two CO metallic phases (M$_{AAB}^1$ and M$_{AAB}^2$) and an insulating one (I$_{AAB}$). The total concentration in the latter is always $4/3$, while in the atomic-limit case it has the sublattice occupation numbers $(2,2,0)$. The M$_{AAB}^1$ phase is in fact a so-called pinball liquid (PL) phase, i.e. the electrons cannot hop through one of the sublattices which takes role of pins, while the honeycomb lattice formed by the other two sublattices is metallic. The phase loses its PL properties as it approaches a non-ordered phase (the dotted line).

The blue region in Fig. \ref{fig:phase-diagram} (subsection \ref{sec:ABB}) encompasses two CO metallic phases (M$_{ABB}^1$ and M$_{ABB}^2$) and an insulating one (I$_{ABB}$, $n=2/3$) which in the atomic-limit case has the sublattice occupation numbers $(2,0,0)$. 

When $U > 0D$, CO metallic phases with the most broken symmetry occur, i.e., all three occupation numbers in the sublattices are different from each other (M$_{ABC}$, white region, Fig. \ref{fig:structures}b, subsection \ref{sec:AABABB}). The two phases can be identified in this region: an $ABB$-like M$_{ABC}$ phase and an $AAB$-like M$_{ABC}$ phase. The degeneracy of the phases is $6$.

\subsection{\textit{AAB} Region}\label{sec:AAB}

\begin{figure}[ht]
    \centering
    \includegraphics[width=1.0\linewidth]{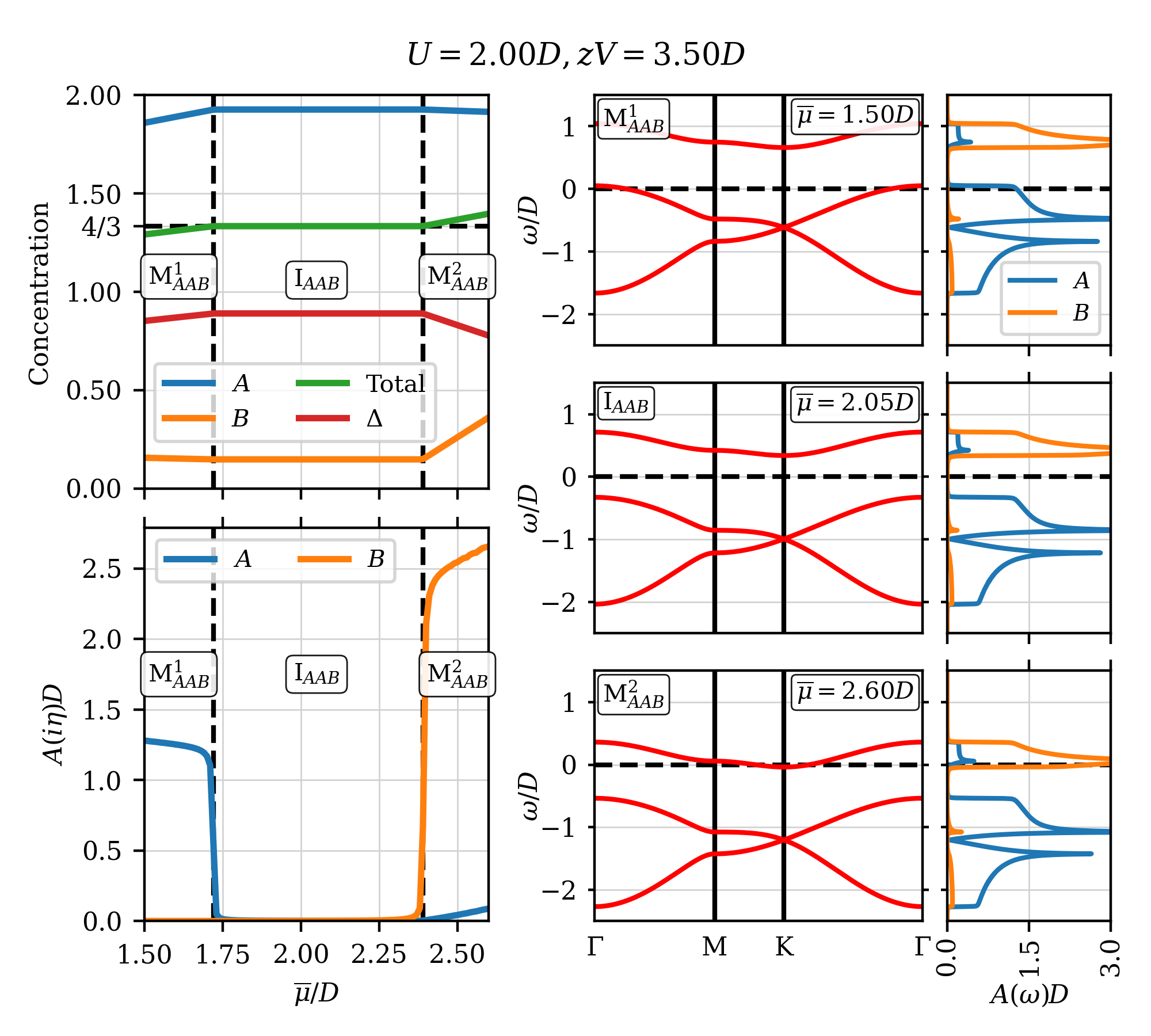}
    \caption{The order parameters in the $AAB$ region (left) and representative band structures of the $AAB$ phases (right). The vertical dashed lines show the locations of phase transitions.}
    \label{fig:AAB}
\end{figure}

Two continuous phase transitions and the band structures of the $AAB$ phases (yellow-region phases in Fig. \ref{fig:phase-diagram}) are shown in Fig. \ref{fig:AAB}. 
For the whole band-structure evolution between $\bar\mu = 1.50D$ and $\bar\mu = 2.60D$ for $U = 2.00D$ and $zV = 3.50D$ see the file \texttt{M1\_U2.00\_V3.50\_AAB.mp4}~\footnote{See Supplemental Material at [URL will be inserted by publisher] with the results for band structure and evolution of various observables as a function of the model parameters.}.
Since sublattices in this region are occupied by two different occupation numbers only, the useful order parameter is a polarization
\begin{equation}
    \Delta = \frac{n_A - n_B}{2}.
\end{equation}
The spectral weight $A_A(\omega) = 2A_1(\omega) = 2A_2(\omega)$ and $A_B(\omega) = A_3(\omega)$.

When not in the proximity of the M$_{AAA}$ or I$_{222}$ phases (see subsection \ref{sec:symbreak}), the \textit{AAB} phases can be described as a separate triangular lattice formed by the sublattice with the occupation $n_B$ (triangular-$B$ lattice), a honeycomb lattice represented by the sublattices with the occupation $n_A$ (honeycomb-$A$ lattice), and a certain amount of hybridization between them (cf. the non-interacting band structures in Fig. \ref{fig:non-interacting}a and \ref{fig:non-interacting}c). The hopping between sites of the triangular-$B$ lattice takes place by means of this hybridization with the honeycomb-$A$ lattice and hance has a modified amplitude: the opposite sign and a smaller absolute value.
As the intersite interaction $V$ increases, the hybridization decreases. This turns the triangular-like band into a dispersionless atomic level, while the lower bands become indistinguishable from the non-interacting honeycomb band structure (Fig. \ref{fig:non-interacting}c) with a shifted Fermi level.
For a band-structure evolution with the increase of $V$ see the file \texttt{M2\_U2.00\_mu2.00\_AAB.mp4} \cite{Note1}. 
The above is valid for all ${AAB}$ phases. 

Noticeably, a position of a Dirac cone at the K-point of the honeycomb-like bands and a position of an extremum at the K-point of the triangular-like band are always fixed at
\begin{equation}\begin{aligned}
    \omega_A = \omega_\text{shift} - \frac{\omega_\Delta}{3}, &&\quad
    \omega_B = \omega_\text{shift} + \frac{2\omega_\Delta}{3},
\end{aligned}\end{equation}
respectively, where the distance between them:
\begin{equation}
    \omega_\Delta = (zV - U)\Delta.
\end{equation}
The other way to write these expressions is
\begin{equation}\label{bandlocation}
    \omega_\alpha = \omega_\text{shift} + (zV-U)\frac{n - n_\alpha}{2}
\end{equation}
(in this subsection, $\omega_A = \omega_1 = \omega_2$ and $\omega_B = \omega_3$).

\begin{figure}[ht]
    \centering
    \includegraphics[width=1.0\linewidth]{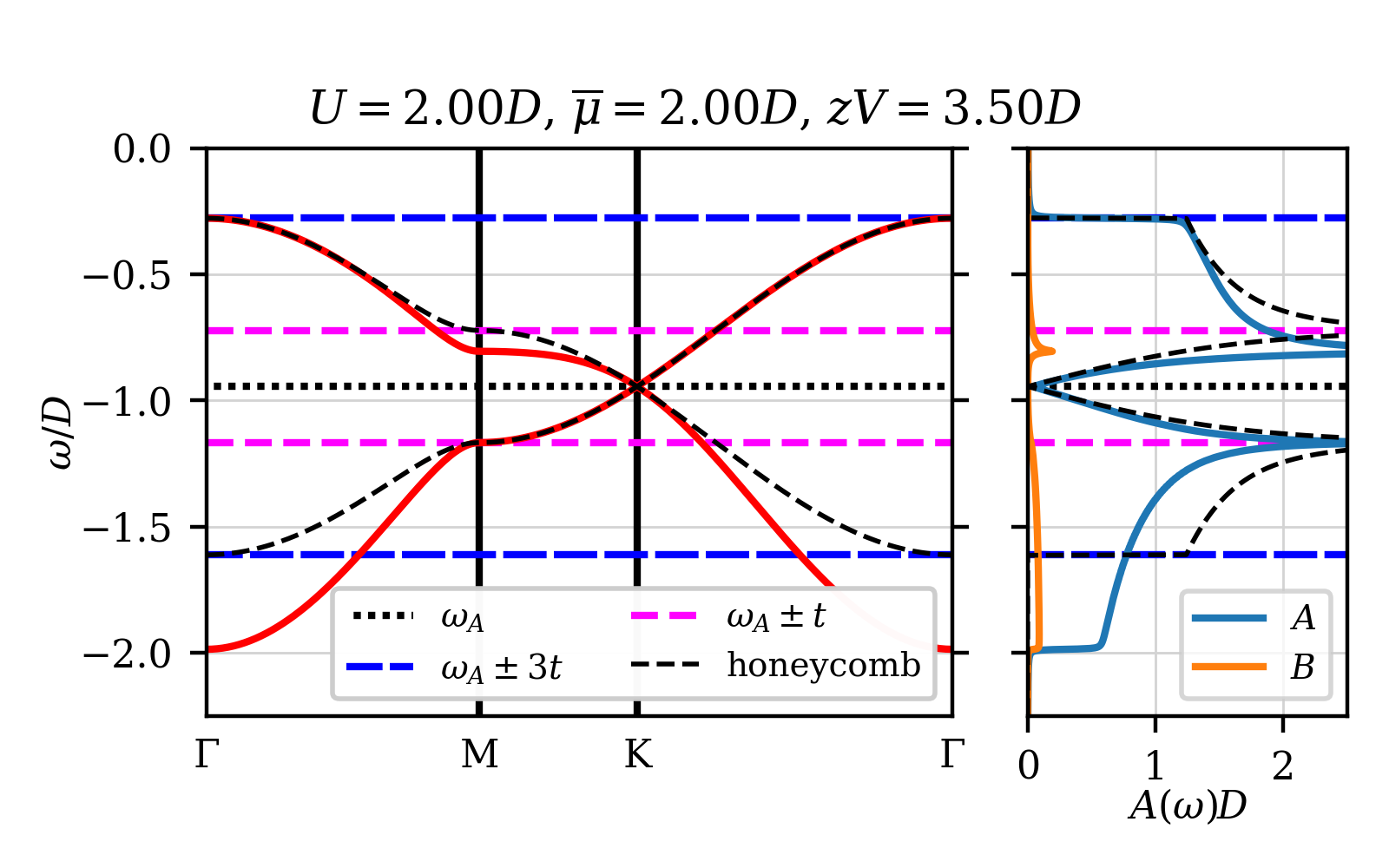}
    \caption{Comparison between the lower bands of an $AAB$-region phase (solid lines) and the non-interacting honeycomb bands (shifted to $\omega_A$).}
    \label{fig:honeycombbands}
\end{figure}

To get more insights on $AAB$ phases, the lower bands (honeycomb-like bands) are shown in Fig. \ref{fig:honeycombbands} together with $\pm 3t$ and $\pm t$ lines where the non-interacting honeycomb band structure has edges and van Hove singularities. It turns out that the upper edge and the lower singularity are always fixed at $\omega_A + 3t$ and $\omega_A - t$, respectively. It justifies the lack of dependency on the interaction $V$ of a left boundary of the I$_{AAB}$ phase in Fig. \ref{fig:phase-diagram}. Consequently, the band gap of the I$_{AAB}$ phase: 
\begin{equation}
    \omega_\text{gap} = \omega_\Delta - 3t = (zV - U)\Delta - 3t
\end{equation}
(asymptotically approaches $zV - U - 3t$ for a large $V$).

On the energy scale the hybridization with the triangular-$B$ lattice is concentrated around the lower edge and the upper singularity of the honeycomb-like bands. It is clear from the non-zero spectral weight of the triangular-$B$ lattice. Locations of these points are not fixed to $-3t$ and $t$, but approach these values in the large-$V$ limit where the hybridization disappears (see the file \texttt{M2\_U2.00\_mu2.00\_AAB.mp4} \cite{Note1}).

Since there is no spectral weight of the triangular-$B$ lattice close to the maximum of the lower bands, the M$_{ABC}^1$ phase is a \emph{pinball liquid}. Close to the non-charge-ordered phase, the upper and lower bands start to merge and the phase is not a PL anymore. This transition is continuous and very smooth (see for example Fig. \ref{fig:var_V}d and \ref{fig:var_V}e \cite{Note1}). To give an approximate boundary of the PL phase, the dotted line in Fig. \ref{fig:phase-diagram} shows the points where the maximum of the lower bands (located at the $\Gamma$-point) and the minimum of the upper band (located at the K-point) have the same energy. From the expressions above it is clear that this line exactly corresponds to $\omega_\Delta = 3t$.

For $U=0D$ or close to $0D$, the I$_{AAB}$-M$_{AAB}^2$ transition is sharp and gets discontinuous for $\bar\mu \gtrsim 3D$ (see for example Fig. S2a \cite{Note1}). 
It can be justified by the proximity of the transition to the fully occupied phase (I$_{222}$). The total concentration in I$_{AAB}$ and I$_{222}$ phases is always $4/3$ and $2$, respectively, while the transition between M$_{AAB}^2$ and I$_{222}$ phases is continuous (subsection \ref{sec:symbreak}). As a consequence the sharp change in occupation numbers occurs at the I$_{AAB}$-M$_{AAB}^2$ transition. When moving from the CO metallic phase, this discontinuous transition corresponds to the release of the potential energy at the expense of the chemical energy.

It is worth noting, since the M$_{AAB}^2$ phase exists for any large values of $V$ (when $\bar\mu$ is also large), and in the large-$V$ limit the hybridization between the honeycomb-$A$ and the triangular-$B$ lattices disappears, it eventually becomes an insulator: the Fermi level is located in the atomic-like peak formed by the sublattice with the occupation $n_B$ (and both $n$ and $n_B$ are not integer!), while the hopping with neighboring sites is restricted. This situation is rather unphysical, and both an inclusion of a next-nearest-neighbor hopping, and an inclusion of the Mott physics can resolve it.

\subsection{\textit{ABB} Region}\label{sec:ABB}
\begin{figure}[ht]
    \centering
    \includegraphics[width=1.0\linewidth]{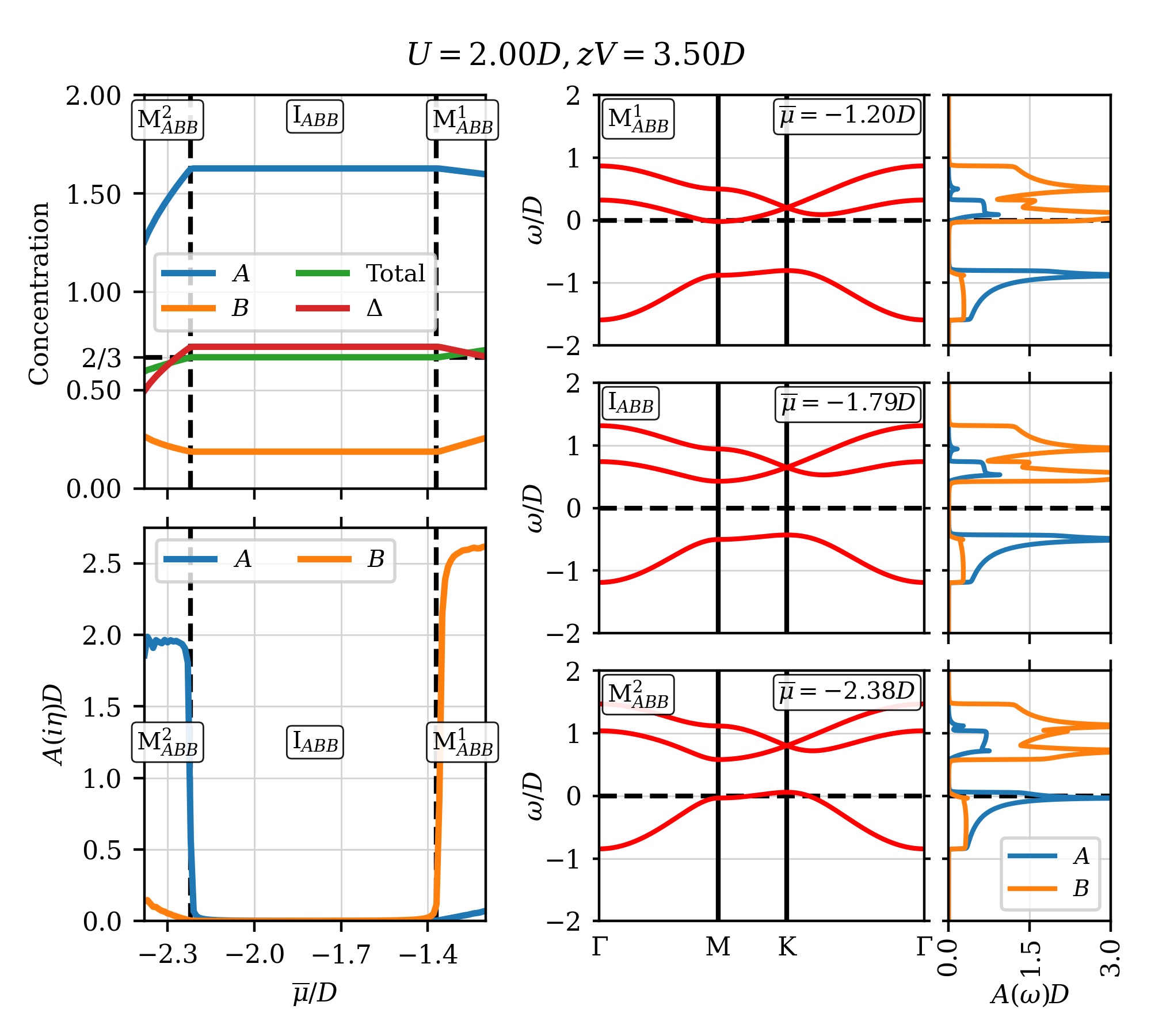}
    \caption{The order parameters in the $ABB$ region (left) and representative band structures the $ABB$ phases (right). The vertical dashed lines show the locations of phase transitions.}
    \label{fig:ABB}
\end{figure}

The band structures of the $ABB$ phases (blue region in Fig. \ref{fig:phase-diagram}) and phase transitions between them are shown in Fig. \ref{fig:ABB}. For the whole band structure evolution during these transitions see also the file \texttt{M3\_U2.00\_V3.50\_ABB.mp4} \cite{Note1}. 
Here, the spectral weight $A_A(\omega) = A_1(\omega)$ and $A_B(\omega)= 2A_2(\omega) = 2A_3(\omega)$.

In the $ABB$ region, honeycomb-like bands of a lattice formed by the sublattices with the occupation $n_B$ (honeycomb-$B$ lattice) are located above a triangular-like band of a lattice formed by the sublattice with the occupation $n_A$ (triangular-$A$). 
It is clear, that the upper bands look less like the honeycomb bands (Fig. \ref{fig:non-interacting}c) for $zV=3.50D$ in comparison to the $AAB$ lower bands for the same $V$. It is justified by the fact that the 1st and the 2nd bands in Fig. \ref{fig:non-interacting}b can be transformed to the honeycomb bands (Fig. \ref{fig:non-interacting}c) with less effort than the 2nd and the 3rd bands. When $V$ increases, the bands change accordingly to form non-interacting honeycomb-like and triangular-like (eventually, an atomic-like) bands while the hybridization between them disappears: see the file \texttt{M4\_U2.00\_mu-1.80\_ABB.mp4} \cite{Note1}. 

Positions of the K-point extremum of the triangular-like band and the Dirac cone of the honeycomb-like bands are still fixed to the energies defined by Eq. (\ref{bandlocation}) ($\omega_A=\omega_1$ and $\omega_B=\omega_2=\omega_3$) or, in terms of the polarization,
\begin{equation}\begin{aligned}
    \omega_A = \omega_\text{shift} - \frac{2\omega_\Delta}{3}, &&\quad
    \omega_B = \omega_\text{shift} + \frac{\omega_\Delta}{3}.
\end{aligned}\end{equation}

Similar to the $AAB$ phases, the upper edge and the lower singularity of the honeycomb-like bands are also fixed to $3t$ and $-t$ in contrast to the lower edge and the upper singularity that approach $-3t$ and $t$ only in the large-$V$ limit when the hybridization disappears. Since, the lower edge is now adjacent to a band gap, the $ABB$ phases are noticeably different from the $AAB$ phases. The right boundary of the I$_{ABB}$ phase in Fig. \ref{fig:phase-diagram} is dependent on $V$. The expression of the lower limit only of the I$_{ABB}$-phase band gap can be written:
\begin{equation}
    \omega_\text{gap} \ge (zV - U)\Delta - 3t.
\end{equation}
When $V$ increases, the $\omega_\text{gap}$ asymptotically approaches the right-side expression.

Since the hybridization between the honeycomb-$B$ and triangular-$A$ lattices is concentrated around the minimum of the honeycomb-like bands, exactly where the Fermi level of the M$_{ABB}^1$ phase is located, the phase is not a PL. 
In the limit of large $V$ the hybridization disappears; hence, in this limit, the M$_{ABB}^1$ phase would eventually become a PL, but the border of such a transition is vague.
Similarly to the M$_{AAB}^2$, the M$_{ABB}^2$ phase in large-$V$ limit is an insulator with non-integer occupations which can be viewed as an artifact of a model or its mean-field treating.

In contrast to the $AAB$ region, the M$_{ABB}^2$ phase does not exist for $U=0D$ or close to $0D$. The discontinuous transition to the non-charge-ordered phase (subsection \ref{sec:symbreak}) takes place directly from I$_{ABB}$ phase.

\subsection{Symmetry Breaking from a Non-Charge-Ordered Phase}\label{sec:symbreak}

Throughout most of the range of $\mu$ the transitions from the metallic non-charge-ordered phase (M$_{AAA}$, red region in Fig. \ref{fig:phase-diagram}) to the CO phases (M$_{ABB}^1$, M$_{ABB}^2$, I$_{ABB}$, M$_{AAB}^1$, M$_{AAB}^2$, I$_{AAB}$) are discontinuous (see for example Fig. \ref{fig:var_V} \cite{Note1}). 
Noticeably, these transitions take place for $zV \gtrsim U + D$. 
The discontinuous symmetry breaking corresponds to the release of intersite-interaction energy at the expense of both kinetic and on-site-interaction energy. Additionally, the chemical energy decreases with transitions to the M$_{ABB}^2$ and I$_{ABB}$ phases, and increases with transitions to the M$_{AAB}^2$ and I$_{ABB}$ phases. The CO phases are metastable inside the M$_{AAA}$ region for a rather small range of parameters: on the $zV$ axis the span of this range is up to $0.1D$ for the most noticeable hysteresis (I$_{ABB}$-M$_{AAA}$ transition for $U=0D$ and large hole doping).

The analysis of band structures shows that the CO phases get unstable when the Fermi level approaches certain valleys or saddle points. Moreover, there are ranges of $\mu$ where the symmetry-breaking transitions to the CO metallic phases are continuous 
(see continuous evolutions in the files \texttt{M*\_SymBreak.mp4} \cite{Note1}).

\subsubsection{\texorpdfstring{M$_{ABB}^1$ and I$_{ABB}$ Phases}{M\textit{ABB}1 and I\textit{ABB} Phases}}
The CO M$_{ABB}^1$ and I$_{ABB}$ phases become unstable when the Fermi level approaches (from above) the maximum of the lower (triangular-like) band located at the K-point (see their phase diagrams in Fig. \ref{fig:ABB} for the reference).

For the continuous transition to the M$_{AAA}$ phase, the maximum of the triangular-like band should merge with the Dirac point of the honeycomb-like bands at the K-point. The Fermi level cannot be above the Dirac point because it destabilizes the M$_{ABB}^1$ phase towards an $AAB$ phase. However, the Fermi level stays between the maximum of the triangular-like band and the Dirac point close to the triple phase-transition point of the M$_{AAA}$, M$_{ABB}^1$, and M$_{AAB}^1$ phases. There, the continuous symmetry breaking to the M$_{ABB}^1$ phase takes place. It is visualized in the file \texttt{M5\_U2.00\_mu-0.85\_ABB\_SymBreak.mp4} \cite{Note1}.

Since the Fermi level appears where all three bands of the non-charge-ordered phase are degenerate at the K-point, the $\omega_\text{shift} = 0t$ around the continuous transition, i.e.
\begin{equation}\label{triplepoint}
    \bar\mu = \left(zV + \frac{U}{2}\right) (n - 1),
\end{equation}
where $n \approx 0.7973$ is the non-interacting triangular-lattice occupation that corresponds to $\omega_\text{F} = 0t$. 
Noticeably, around the triple point the $zV \approx U+D$ which brings us to the expression:
\begin{equation}
    \frac{\bar\mu}{D} \approx 0.10 - 0.30\frac{zV}{D}
\end{equation}
for the triple point.

\subsubsection{\texorpdfstring{M$_{AAB}^1$ Phase}{M\textit{AAB}1 Phase}}

Similarly, the M$_{AAB}^1$ phase (but not I$_{AAB}$) becomes unstable towards the M$_{AAA}$ phase when the Fermi level approaches (from below) the minimum of the upper (triangular-like) band at the K-point (see its phase diagram in Fig. \ref{fig:AAB} for the reference). The same as for the M$_{ABB}^1$ phase, the continuous phase transition (merging the extremum of the triangular-like band and the Dirac point at the K-point) happens around the same triple point of the phase diagram where Eq. (\ref{triplepoint}) is satisfied with the same $n$. 
The continuous transition is visualized in the file \texttt{M6\_U2.00\_mu-0.80\_AAB\_SymBreak.mp4} \cite{Note1}.

\subsubsection{\texorpdfstring{M$_{ABB}^2$ Phase}{M\textit{ABB}2 Phase}}

For $U$ close to $0D$, there is no transition between the M$_{AAA}$ and M$_{ABB}^2$ phases. For a larger $U$, the CO phase becomes unstable towards the M$_{AAA}$ phase in two ways: 1) when the Fermi level approaches (from above) the singularity (saddle point) of the lower (triangular-like) band at the M-point; 2) when the Fermi level approaches (from below) the minimum of the upper (honeycomb-like) bands which in this range of parameters is at the M-point. 

The continuous transition between the M$_{AAA}$ and M$_{ABB}^2$ phase happens in the small range of $\mu$ when the Fermi level stays in between of the mentioned saddle point and minimum at the M-point. It is visualized in the file \texttt{M7\_U2.00\_mu-1.80\_ABB\_SymBreak.mp4} \cite{Note1}.
Since the Fermi level appears where the two bands of the M$_{AAA}$ phase are degenerate at the M-point, the $\omega_\text{shift} = t$ around the continuous transition, i.e.
\begin{equation}
    \bar\mu + t = \left(zV + \frac{U}{2}\right) (n - 1),
\end{equation}
where $n \approx 0.6078$ is the non-interacting triangular-lattice occupation when $\omega_\text{F} = -t$.

\subsubsection{\texorpdfstring{M$_{AAB}^2$ and I$_{AAB}$ Phases}{M\textit{AAB}2 and I\textit{AAB} Phases}}

The M$_{AAB}^2$ and I$_{AAB}$ phases become unstable towards the M$_{AAA}$ phase when the Fermi level approaches (from above) the maximum of the lower (honeycomb-like) bands at the $\Gamma$-point. 

In contrast to the M$_{ABB}^2$, the M$_{AAB}^2$ phase cannot continuously transform to the M$_{AAA}$ phase but has an infinite range of $\mu$ (particularly, $\bar\mu \gtrsim 3(U/2+D)$) where it continuously transforms to the fully occupied phase (I$_{222}$). The continuous transition is visualized in the file \texttt{M8\_U0.00\_mu3.10\_AAB\_SymBreak.mp4} \cite{Note1}.

At this continuous transition the $\omega_\text{shift} = -3t$ with $n=2$ which corresponds to the top of the non-interacting triangular-lattice bands:
\begin{equation}
    \bar\mu - 3t = zV + \frac{U}{2}.
\end{equation}

\subsection{\textit{AAB}-\textit{ABB} Transition and \textit{ABC} Region}\label{sec:AABABB}

The transition between the M$_{ABB}^1$ and M$_{AAB}^1$ phases (between the yellow and blue regions) is discontinuous (see for example Fig. \ref{fig:var_mu}a \cite{Note1}). When moving from the $ABB$ phase, it happens with the release of chemical energy at the expense of intersite-interaction energy.

For $U>0D$ an intermediate region of M$_{ABC}$ phases ($n_1=n_A$, $n_2=n_B$, $n_3=n_C$, $n_A>n_B>n_C$, white region) opens between the $ABB$ and $AAB$ regions (see for example Fig. S3d \cite{Note1}). For the most range of parameters the transition between the $ABC$ region and the M$_{ABB}^1$ phase on one side and the M$_{AAB}^1$ phase on the other side is continuous. Note that the less-symmetry-broken phases ($ABB$ and $AAB$) can have the self-consistent solution of the Hamiltonian (\ref{hamiltonian}) inside the $ABC$ region despite other indicators that the phase transition is continuous.

The $ABC$ region can be characterized by identifying two phases: the $ABB$-like M$_{ABC}$ phase and the $AAB$-like M$_{ABC}$ phase with the discontinuous transition between them. Moreover, there is a range of $V$ where the $AAB$-like M$_{ABC}$ phase discontinuously transforms directly into the M$_{ABB}^1$ phase (see Fig. \ref{fig:phase-diagram}). The same as for the M$_{ABB}^1$-M$_{AAB}^1$ transition, these discontinuous transitions take place with the release of chemical energy at the expense of intersite-interaction energy when moving from the $ABB$-like side.

There is a range of parameters where the $ABC$ phase can be viewed as both $ABB$-like and $AAB$-like M$_{ABC}$ phase (in Fig. \ref{fig:phase-diagram} the region around $zV=4.4D$ for $U=2D$). The visualization of two continuous transitions between the $ABB$, $ABC$, and $AAB$ regions for $U=2D$ and $V=4.4D$ is presented in the file \texttt{M9\_U2.00\_V4.40 \_ABB-ABC-AAB.mp4} \cite{Note1}.

The band structure of the M$_{ABC}$ phase consists of $3$ bands: a lower triangular-like which spectral weight primarily comes from the $A$-sublattice, an upper triangular-like which spectral weight primarily comes from the $C$-sublattice, and an intermediate band which spectral weight primarily comes from the $B$-sublattice. The Fermi level is located inside the intermediate band. All three bands have an extremum at the K-point that is fixed to the energy defined from Eq. (\ref{bandlocation}) ($\omega_A=\omega_1$, $\omega_B=\omega_2$, and $\omega_C=\omega_3$). In the large-$V$ limit the three bands are rather three atomic levels: a fully occupied level, a fully unoccupied level, and a level where the Fermi level is located.

\subsection{Correspondence to the Atomic-Limit Phase Diagram}\label{atlim}

For a large $V$ the atomic-limit ($D=0$) solution  of the triangular-lattice EHM (exact for the ground state) predicts $7$ phases that are denoted according to the three occupation numbers: 000, 100, 200, 210, 220, 221, and 222 \cite{kapcia2021}. The phase transitions take place at the following lines:
\begin{eqnarray}\label{atomic-limit}
    \bar\mu &= &\mp (zV + U/2),\ \text{for 000-100 and 222-221}, \nonumber\\
    \bar\mu &= &\mp (zV - U/2),\ \text{for 100-200  and 221-220}, \quad \\
    \bar\mu & = &\mp U/2, \quad \ \ \qquad \text{for 200-210  and 220-210}, \nonumber
\end{eqnarray}
where the '$\mp$' corresponds to the first or second boundary mentioned above, respectively.
It is clear from the formulas that the phases 100, 210 and 221 do not exist for $U=0D$.

There is a correspondence between the atomic-limit and mean-field solutions at a large $V$. Besides the obvious correspondence for the fully-unoccupied and fully-occupied phases, the atomic-limit 100 phase can be associated with the mean-field M$_{ABB}^2$ phase, the 200 phase with the I$_{ABB}$ and M$_{ABB}^1$ phases together, the 210 phase with M$_{ABC}$ phase, and correspondingly for the other side of the phase diagram.

Thus, the disappearance of the M$_{ABC}$ and M$_{ABB}^2$ phases for $U=0D$ could be predicted from the atomic-limit solution. Despite this correspondence, the mean-field phase M$_{AAB}^2$ exists even for $U=0D$ and large $V$, which makes it an unexpected effect of non-zero itineracy.

To compare the atomic-limit and the mean-field solutions, one can put $U=2D$ in Eqs. (\ref{atomic-limit}) (despite the fact that $D=0$ in the atomic limit):
\begin{eqnarray}
    \bar\mu & = &\mp (zV + D), \ \text{for 000-100 (`--') and 222-221 (`+')}, \nonumber\\
    \bar\mu & = &\mp (zV - D), \ \text{for 100-200 (`--') and 221-220 (`+')}, \nonumber\\
    \bar\mu & = &\mp D, \qquad \ \ \quad \text{for 200-210 (`--') and 220-210 (`+')}. \nonumber
\end{eqnarray}
The mean-field boundaries of M$_{ABC}$ phase at large $V$ and $U=2D$ indeed slowly approach $\bar\mu = \mp D$ (Fig. \ref{fig:phase-diagram}). The same is valid for the boundaries of the I$_{000}$ and I$_{222}$ phases since the terms $-6t$ and $3t$ become negligible in comparison to the large $zV$. The left boundary of the I$_{ABB}$ phase coincide well with the left boundary of the atomic-limit 200 phase even at the small values of $V$, and the same is valid for the 220 and I$_{AAB}$ phases. 

It is reasonable to expect that primarily those phases that are associated with non-zero $U$ (M$_{ABC}$, M$_{ABB}^2$, M$_{AAB}^2$) will be modified when taking into account the Mott physics (local correlations). Note also, that in the atomic limit there are also 111 (Mott insulator), 110, and 211 phases for small $V$ (particularly, for $zV \leq 2U$).

\section{Summary}\label{summary}

The full zero-temperature phase diagram of the extended Hubbard model on the triangular lattice in the all-encompassing range of chemical potential values and repulsive onsite and nearest-neighbor electron interactions is presented and analyzed. Despite the limitations of the mean-field approximation, restriction to the $\sqrt{3}\times\sqrt{3}$ supercells and the absence of magnetic order, the large variety of features are found. Those are the diverse and numerous phase transitions, including the continuous and discontinuous symmetry breakings; the pinball liquid phase; the strong particle-hole asymmetry manifested in every phase transition with a charge-ordered phase. 
Both the found phases and the found phase transitions are accompanied by the extensive band-structure analysis that provided clarity and the understanding of the phase diagram. 

It is worth noting, that the quarter-filling for electrons ($n=0.5$) and holes ($n=1.5$) that is common for some organic conductors occurs in the M$_{AAA}$, M$_{ABB}^2$, and M$_{AAB}^2$ phases, which are not pinball-liquid phases within the MFA.
Nevertheless, for a large enough $U$, we expect that the M$_{ABB}^2$, M$_{AAB}^2$, and M$_{ABC}$ phases should experience prominent Mott localization at one of the sublattices which is consistent with the correspondence to the exact atomic-limit results. This localization may contribute to formation of a pinball liquid \cite{merino2013}. Moreover, for a region $zV \lesssim 2U$ we except the appearance of a non-ordered Mott insulator and the charge-ordered phases that correspond to the 110 and 211 phases of an atomic-limit phase diagram \cite{kapcia2021}. Additionally, when considered, magnetic orders might be found particularly for $zV < U/2$ \cite{Micnas1990,Georges1996}, but detailed analysis of the magnetism in the system is out of the scope of this work.

It is evident that the future investigation beyond the mean-field approximation can benefit from our analysis.
Additionally, our research can easily be expanded to include next-nearest-neighbor interactions and finite temperatures. The more work is also required to include magnetic order and to consider the orderings that require larger supercells. 

\begin{acknowledgments}
K.J.K. thanks the Polish National Agency for Academic Exchange for funding in the frame of the National Component of the Mieczysław Bekker program (2020 edition) (BPN/BKK/2022/1/00011).
\end{acknowledgments}

\bibliography{refs}

\clearpage

\onecolumngrid

\begin{center}
  \textbf{\Large Supplemental Material}\\[.2cm]
  \textbf{\large Particle-Hole Asymmetry and Pinball Liquid in a Triangular-Lattice Extended Hubbard Model within Mean-Field Approximation}\\[.2cm]
  Aleksey Alekseev,$^{1}$ Agnieszka Cichy,$^{1,2}$ Konrad Jerzy Kapcia,$^{1}$ \\ [.2cm]
  {\itshape
  	$^{1}$\mbox{Institute of Spintronics and Quantum Information, Faculty of Physics and Astronomy}, \mbox{Adam Mickiewicz University in Pozna\'n}, 
    Uniwersytetu Pozna\'{n}skiego 2, PL-61614 Pozna\'{n}, Poland\\
	$^{2}$Institut f\"{u}r Physik, Johannes Gutenberg-Universit\"{a}t Mainz, Staudingerweg 9, D-55099 Mainz, Germany\\
	}
(Dated: \today)
\\[1cm]
\end{center}

\setcounter{equation}{0}
\renewcommand{\theequation}{S\arabic{equation}}
\setcounter{figure}{0}
\renewcommand{\thefigure}{S\arabic{figure}}
\setcounter{section}{0}
\renewcommand{\thesection}{S\arabic{section}}
\setcounter{table}{0}
\renewcommand{\thetable}{S\arabic{table}}
\setcounter{page}{1}

In this Supplemental Material we present additional results which clarify the findings included in the main text and extend the discussion, in particular, concerning:
\begin{itemize}
    \item[(i)]{the dependencies of various observables near phase transitions to supplement the discussion of the phase diagram  (see Sec. \ref{sec:SMproperties} including the figures and multimedia description)}
    \item[(ii)]{the full list of all transitions found in the system in the used approach (Sec. \ref{sec:SMlist}).}
\end{itemize}

\section{The Dependencies of Order parameters and Grand-potential Contributions}\label{sec:SMproperties}

\begin{figure}[ht]
    \centering
    \includegraphics[width=0.85\linewidth]{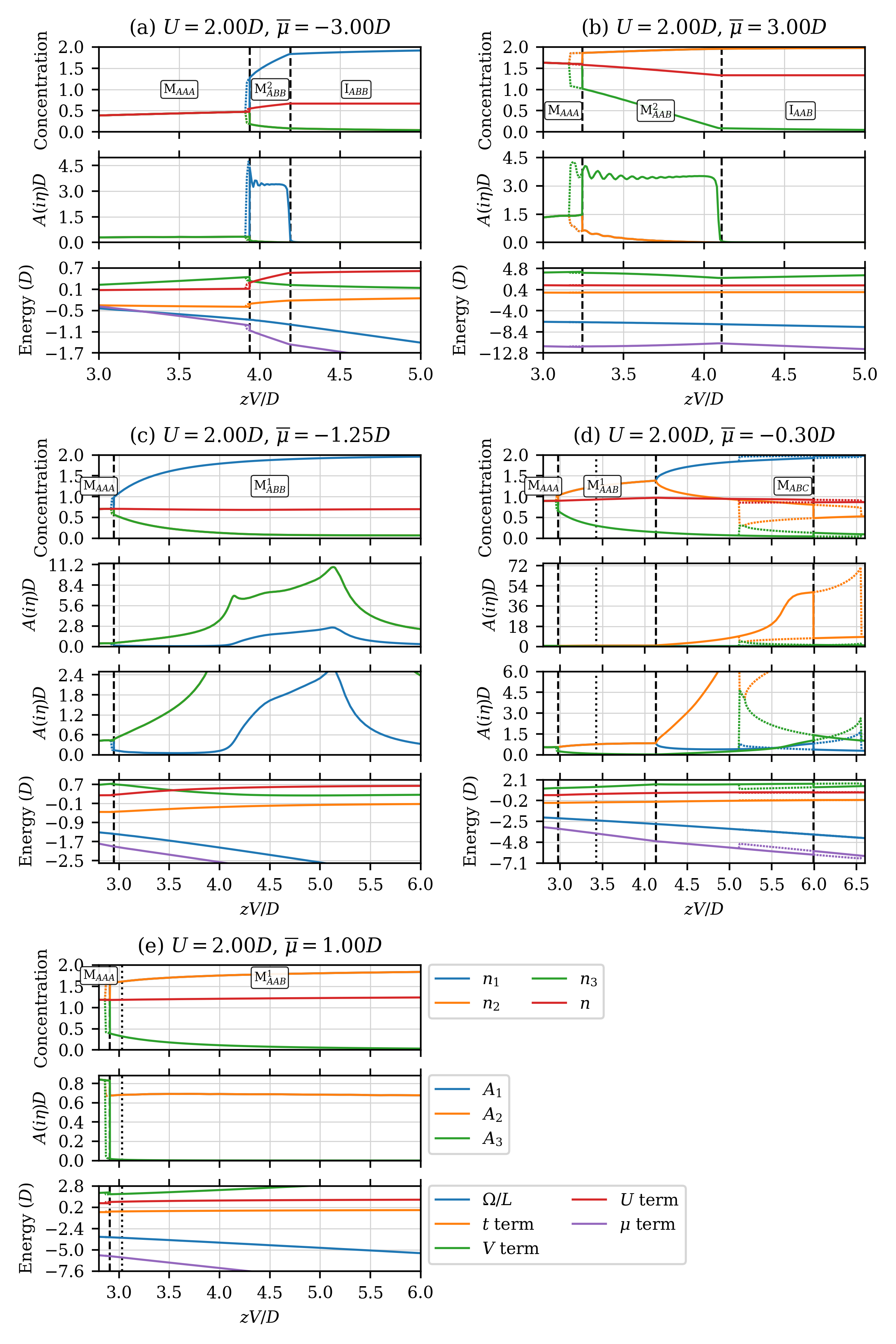}
    \caption{The order parameters and the grand-potential contributions as a function of $V$ for $U= 2.00D$ and representative values of $\bar{\mu}$ (as labeled).
    The black dashed vertical lines show the location of phase transitions, the black dotted vertical line corresponds to the dotted line in Fig. 2 in the main text, where the M$_{AAB}^1$ phase is not a PL anymore.}
    \label{fig:var_V}
\end{figure}

\begin{figure}[ht]
    \centering
    \includegraphics[width=1\linewidth]{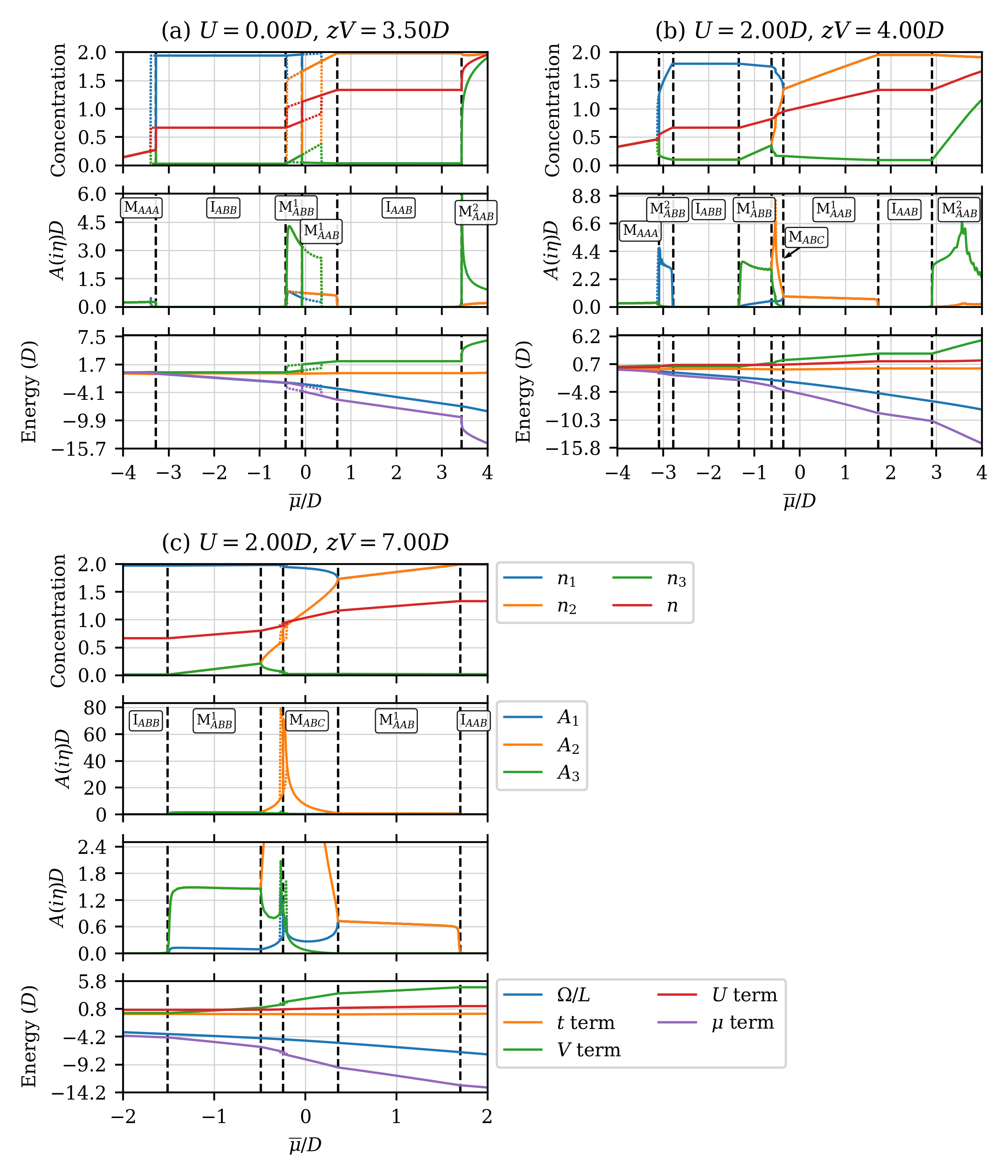}
    \caption{The order parameters and the grand-potential contributions as a function of $\bar{\mu}$ for representative values of $U$ and $V$ (as labeled). 
    The black dashed vertical lines show the location of phase transitions.}
    \label{fig:var_mu}
\end{figure}

\begin{figure}[ht]
    \centering
    \includegraphics[width=1\linewidth]{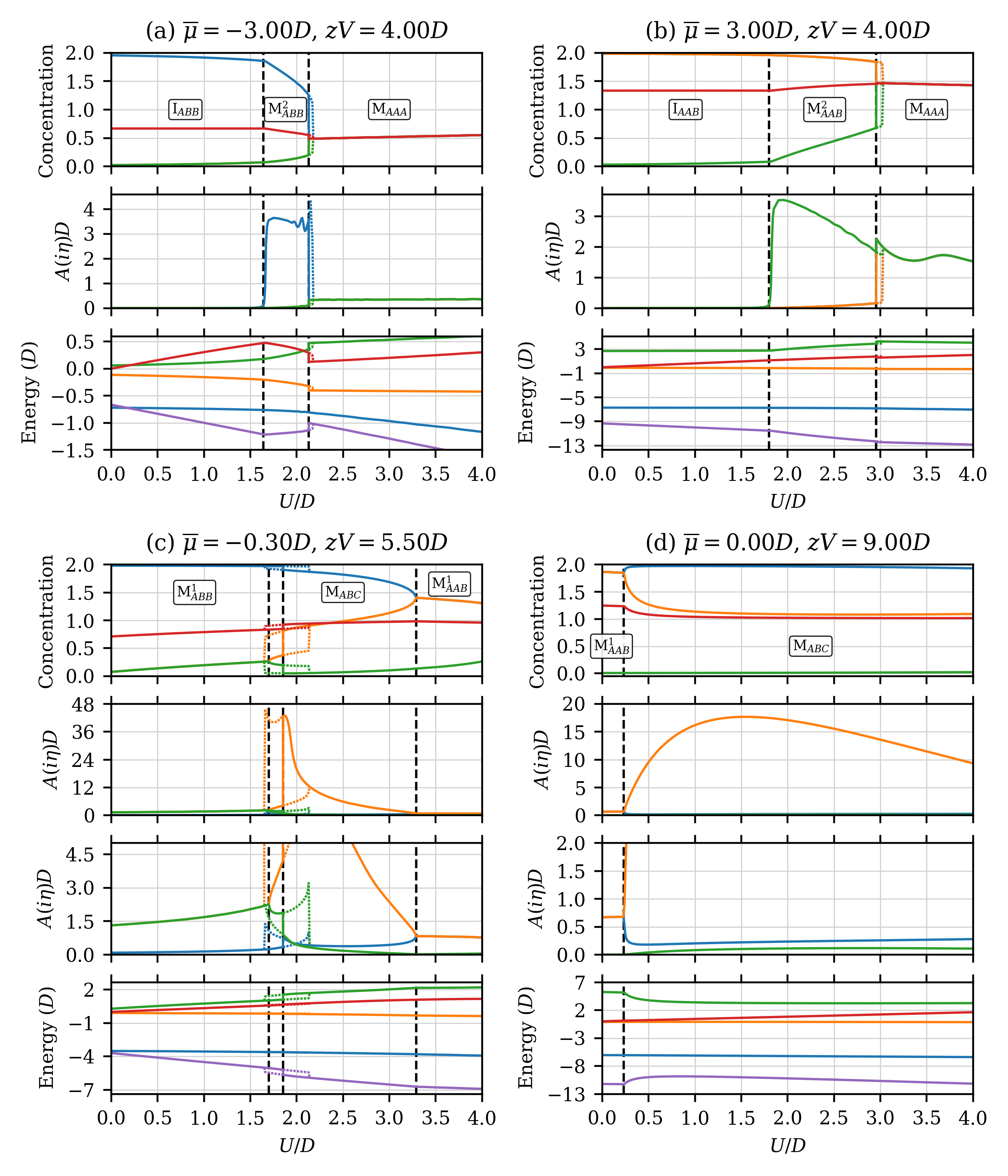}
    \caption{The order parameters and the grand-potential contributions as a function of $U$ for representative values of $\bar{\mu}$ and $V$ (as labeled). 
    The black dashed vertical lines show the location of phase transitions. 
    The legend is the same as in Figs. \ref{fig:var_V} and \ref{fig:var_mu}.}
    \label{fig:var_U}
\end{figure}

Figures \ref{fig:var_V}, \ref{fig:var_mu}, and \ref{fig:var_U} show the order parameters and the contributions in the grand potential as a function of $V$, $\bar{\mu}$, and $U$, respectively. 
The metastability of the phases (hysteresis) is also shown with dotted lines. 
In some cases the spectral weight at the Fermi level ($A(i\eta)$) is shown on two different scales due to its proximity to a spectral-weight singularity in the $M_{ABB}^1$ and $M_{ABC}$ phases.
In particular, the figures contain the results for the following model parameters:
\begin{itemize}
    \item{Fig. \ref{fig:var_V} shows $V$-dependencies for $U=2.00 D$ and various values of $\bar{\mu}$: 
    (a)  $\bar{\mu} =- 3.00 D$, 
    (b) $\bar{\mu} = 3.00 D$, 
    (c)  $\bar{\mu} =- 1.25 D$,
    (d) $\bar{\mu} =- 0.30 D$, and 
    (e) $\bar{\mu} = 1.00 D$;}
    \item{Fig. \ref{fig:var_mu} shows $\bar\mu$-dependencies for 
    (a) $U=0.00 D$ and  $zV = 3.50 D$,
    (b) $U=0.00 D$ and  $zV = 4.00 D$, and
    (c) $U=2.00 D$ and  $zV = 7.00 D$;}
    \item{Fig. \ref{fig:var_U} shows $U$-dependencies for
    (a) $\bar{\mu} =- 3.00 D$ and  $zV = 4.00 D$,
    (b) $\bar{\mu} = 3.00 D$ and  $zV = 4.00 D$,
    (c) $\bar{\mu} =- 0.30 D$ and  $zV = 5.50 D$, and
    (d) $\bar{\mu} = 0.00 D$ and  $zV = 9.00 D$.}
\end{itemize}

In addition, we present nine multimedia movies that show the evolution of the band structure and density of states:
\begin{enumerate}
    \item{\texttt{M1\_U2.00\_V3.50\_AAB.mp4}\\ 
    ---  evolution between $\bar\mu = 1.50D$ and $\bar\mu = 2.60D$ for $U = 2.00D$ and $zV = 3.50D$ (transitions M$_{AAB}^1$-I$_{AAB}$ and I$_{AAB}$-M$_{AAB}^2$)}
    \item{\texttt{M2\_U2.00\_mu2.00\_AAB.mp4}\\ 
    ---  evolution between $zV = 3.50D$ and $zV = 9.00D$ for $U = 2.00D$ and $\bar{\mu} = 2.00D$}
    \item{\texttt{M3\_U2.00\_V3.50\_ABB.mp4}\\
    ---  evolution between $\bar\mu = -2.38D$ and $\bar\mu = -1.20D$ for $U = 2.00D$ and $zV = 3.50D$ (transitions M$_{ABB}^2$-I$_{ABB}$ and I$_{ABB}$-M$_{ABB}^1$)}
    \item{\texttt{M4\_U2.00\_mu-1.80\_ABB.mp4}\\
    --- evolution between $zV = 3.50D$ and $zV = 9.00D$ for $U = 2.00D$ and $\bar{\mu} = -1.80D$}
    \item{\texttt{M5\_U2.00\_mu-0.85\_ABB\_SymBreak.mp4}\\
    --- evolution between $zV = 2.95D$ and $zV = 3.45D$ for $U = 2.00D$ and $\bar{\mu} = -0.85D$ (transition M$_{AAA}$-M$_{ABB}^1$)}
    \item{\texttt{M6\_U2.00\_mu-0.80\_AAB\_SymBreak.mp4}\\
    --- evolution between $zV = 2.95D$ and $zV = 3.45D$ for $U = 2.00D$ and $\bar{\mu} = -0.80D$ (transition M$_{AAA}$-M$_{AAB}^1$)}
    \item{\texttt{M7\_U2.00\_mu-1.80\_ABB\_SymBreak.mp4}\\
    --- evolution between $zV = 2.95D$ and $zV = 3.45D$ for $U = 2.00D$ and $\bar{\mu} = -1.80D$ (transitions M$_{AAA}$-M$_{ABB}^2$ and M$_{ABB}^2$-I$_{ABB}$)}
    \item{\texttt{M8\_U0.00\_mu3.10\_AAB\_SymBreak.mp4}\\
    --- evolution between $zV = 2.35D$ and $zV = 2.85D$ for $U = 0.00D$ and $\bar{\mu} = 3.10D$ (transition I$_{222}$-M$_{AAB}^2$)}
    \item{\texttt{M9\_U2.00\_V4.40\_ABB-ABC-AAB.mp4}\\
    --- evolution between $\bar{\mu} = -0.70D$ and $\bar{\mu} = -0.10D$ for $U = 2.00D$ and $zV = 4.40D$ (transitions M$_{ABB}^1$-M$_{ABC}$ and M$_{ABC}$-M$_{AAB}^1$)}
\end{enumerate}

\section{Transitions Found in the System}\label{sec:SMlist}

The list of 24 phase transitions:
\renewcommand{\labelenumii}{\arabic{enumi}.\arabic{enumii}}
\begin{enumerate}
    \item $ABB$-metal-insulator transitions (blue region):
    \begin{enumerate}
        \item M$_{ABB}^1$ to I$_{ABB}$ (continuous);
        \item M$_{ABB}^2$ to I$_{ABB}$ (continuous);
    \end{enumerate}
    
    \item $AAB$-metal-insulator transitions (yellow region):
    \begin{enumerate}
        \item M$_{AAB}^1$ to I$_{AAB}$ (continuous);
        \item M$_{AAB}^2$ to I$_{AAB}$ (continuous);
        \item M$_{AAB}^2$ to I$_{AAB}$ (discontinuous), for large $\bar\mu$ but small $U$;
    \end{enumerate}

    \item M$_{AAA}$ to $ABB$ transitions (red to blue):
    \begin{enumerate}
        \item M$_{AAA}$ to M$_{ABB}^1$ (discontinuous);
        \item M$_{AAA}$ to M$_{ABB}^1$ (continuous), a small region in the very proximity of M$_{ABB}^1$-M$_{AAB}^1$ transition;
        \item M$_{AAA}$ to M$_{ABB}^2$ (discontinuous), not for small $U$;
        \item M$_{AAA}$ to M$_{ABB}^2$ (continuous), a small region, not for small $U$;
        \item M$_{AAA}$ to I$_{ABB}$ (discontinuous);
    \end{enumerate}

    \item M$_{AAA}$ to $AAB$ transitions (red to yellow):
    \begin{enumerate}
        \item M$_{AAA}$ to M$_{AAB}^1$ (discontinuous);
        \item M$_{AAA}$ to M$_{AAB}^1$ (continuous), a small region in the very proximity of M$_{ABB}^1$-M$_{AAB}^1$ transition;
        \item M$_{AAA}$ to M$_{AAB}^2$ (discontinuous);
        \item M$_{AAA}$ to I$_{AAB}$ (discontinuous);
    \end{enumerate}

    \item M$_{ABB}^1$-M$_{AAB}^1$ transition (blue to yellow) (discontinuous);

    \item $ABC$ region (blue-white-yellow):
    \begin{enumerate}
        \item M$_{ABC}$ to M$_{ABB}^1$ (continuous);
        \item M$_{ABC}$ to M$_{AAB}^1$ (continuous);
        \item inside the M$_{ABC}$ the discontinuous transition from ${ABB}$-like M$_{ABC}$ phase to ${AAB}$-like M$_{ABC}$ phase. As shown on phase diagram, there is a region where this transition disappears (the line is terminated)
        \item directly from ${AAB}$-like M$_{ABC}$ phase to M$_{ABB}^1$ phase (discontinuous);
    \end{enumerate}
    
    \item The M$_{AAB}^1$ phase is not a PL as it gets closer to M$_{ABB}^1$-M$_{AAA}$-M$_{AAB}^1$ triple point. The approximate border between M$_{AAB}^1$-PL and M$_{AAB}^1$-non-PL is the dotted line;

    \item In the limit of large $V$, the M$_{ABB}^1$ phase becomes a PL because the hybridization between triangular and honeycomb lattices disappears, but the border of this phase is hard to define;
    
    \item Additionally, there is one transition from M$_{AAA}$ to the fully unoccupied phase, and two transitions to the fully occupied phase: from M$_{AAA}$ and directly from M$_{AAB}^2$ phase.
\end{enumerate}

\end{document}